\documentclass[useAMS,usenatbib]{mn2e}
\usepackage{graphicx}
\usepackage{textcomp}
\usepackage{amssymb}\usepackage{hyperref}

 \usepackage{times}

\title[IMBH in AGN disks II.]{Intermediate mass black holes in AGN disks II. Model predictions and observational constraints}

\author[B. McKernan, K.E.S. Ford, B. Kocsis, W. Lyra, L.M.Winter]{B. McKernan$^{1,2,3,4}$\thanks{E-mail:bmckernan at amnh.org (BMcK)}, K.E.S. Ford$^{1,2,3,4}$, B.Kocsis$^{5,6}$,W.Lyra$^{7,8,10}$, L.M.Winter$^{9}$ \\
$^{1}$Department of Science, Borough of Manhattan Community College, City University of New York, New York, NY 10007, USA\\
$^{2}$Department of Astrophysics, American Museum of Natural History, New York, NY 10024, USA\\
$^{3}$Graduate Center, City University of New York, 365 5th Avenue, New York, NY 10016, USA\\
$^{4}$Kavli Institute for Theoretical Physics, UC Santa Barbara, CA 93106, USA\\
$^{5}$Harvard-Smithsonian Center for Astrophysics, 60 Garden St., Cambridge, MA 02138, USA\\
$^{6}$Institute for Advanced Study, Einstein Drive, Princeton, NJ 08540\\
$^{7}$Jet Propulsion Laboratory, California Institute of Technology, 4800 Oak Grove Dr., Pasadena, CA 91109, USA\\
$^{8}$California Institute of Technology, Division of Geological \& Planetary Sciences,  1200 E. California Blvd., Pasadena, CA 91125, USA\\
$^{9}$Center for Astrophysics \& Space Astronomy, University of Colorado, Boulder, CO 80303, USA\\
$^{10}$Sagan Fellow\\
\\
}
\begin{document}

\date{Accepted. Received; in original form}

\pagerange{\pageref{firstpage}--\pageref{lastpage}} \pubyear{2008}

\maketitle

\label{firstpage}

\begin{abstract}
If intermediate mass black holes (IMBHs) grow efficiently in gas disks around supermassive black holes, their host active galactic nucleus (AGN) disks should exhibit myriad observational signatures. Gap-opening IMBHs in AGN disks can exhibit spectral features and variability analagous to gapped protoplanetary disks. A gap-opening IMBH in the  innermost disk imprints ripples and oscillations on the broad Fe~K$\alpha$ line which may be detectable with future X-ray missions. A non-gap-opening IMBH will accrete and produce a soft X-ray excess relative to continuum emission. An IMBH on a retrograde orbit in an AGN disk will not open a gap and will generate soft X-rays from a bow-shock 'headwind'. Accreting IMBH in a large cavity can generate ULX-like X-ray luminosities and LINER-like optical line ratios from local ionized gas. We propose that many LINERs house a weakly accreting MBH binary in a large central disk cavity and will be luminous sources of gravitational waves (GW). IMBHs in galactic nuclei may also be detected via intermittent observational signatures including: UV/X-ray flares due to tidal disruption events, asymmetric X-ray intensity distributions as revealed by AGN transits, quasi-periodic oscillations and underluminous Type Ia supernovae. GW emitted during IMBH inspiral and collisions may be detected with eLISA and LIGO, particularly from LINERs. We summarize observational signatures and compare to current data where possible or suggest future observations.
\end{abstract}

\begin{keywords}
galaxies: active --
(stars:) binaries:close -- planets-disc interactions -- protoplanetary discs -- 
emission: accretion 
\end{keywords}

\section{Introduction}
\label{sec:intro}
There is overwhelming observational evidence for supermassive black holes (SMBH;$>10^{6}M_{\odot}$) in the centers of galaxies \citep{b4} and stellar mass black holes ($<20M_{\odot}$) in our own Galaxy \citep{b2}. However, there is only fragmentary evidence for intermediate mass black holes (IMBHs, see \citealt{davis11} for the best candidate to date). IMBHs are thus a key missing component of our Universe. The standard model of IMBH production is in clusters \citep{b81}, however there are currently no undisputed cases of IMBHs in globular clusters \citep{b11}; IMBHs are hard to find.

In \citet{b93} (Paper I), we described a model for the production and growth of 
IMBH seeds in disks around supermassive black holes. Our model grows IMBH seeds at  super-Eddington rates in disks in active galactic nuclei (AGN). IMBH growth occurs both via collision of stars and compact objects at low relative velocity in disks (core accretion) and via gas accretion as the IMBH migrates within the disk. Our model is analagous to the growth of giant planets in protoplanetary disks 
\citep[e.g.][]{b89,b20} and grows IMBHs more efficiently than the standard model \citep{b81}. In this paper we outline the wide range of predicted observables that can reveal IMBHs in galactic nuclei throughout the local Universe.

In Section~\S\ref{sec:observations} we discuss conditions in the AGN disk under which 
IMBHs can open a gap. In Section~\S\ref{sec:gap} we discuss observational consequences of gaps in the outer AGN disk. We draw a parallel between observational signatures of gapped protoplanetary disks and gaps carved out
by IMBHs in AGN disks. In section~\S\ref{sec:feka} we outline the effect of a gap-opening IMBH in the inner AGN disk on the broad component of the FeK$\alpha$ line, which yields signatures that allow us to follow the final stage of mergers and provides advance warning of gravitational wave outbursts. In Section~\S\ref{sec:disk_nogap} we discuss the signatures of accreting IMBHs in AGN disks. In Section~\S\ref{sec:indep} we discuss occasional signatures of IMBHs in 
galactic nuclei. In section~\S\ref{sec:gw} we
discuss gravitational wave signatures of our model potentially detectable with LIGO and LISA. 

\section{Gap-opening in AGN disks}
\label{sec:observations}
Gap opening in any accretion disk depends on the ratio of satellite mass ($M_{2}$) to primary mass ($M_{1}$) , disk viscosity and the disk aspect ratio. A simple 
estimate of the critical mass ratio ($q=M_{2}/M_{1}$), above which the IMBH will open a gap in the disk, is given by \citep{linpap86}
\begin{equation}
q \approx \left( \frac{27\pi}{8}\right)^{1/2} \left( \frac{H}{r} 
\right)^{5/2} \alpha^{1/2}
\label{eq:qcrit}
\end{equation}
where ($H/r$) is the disk thickness and $\alpha$ is the viscosity parameter \citep{b66}. 
The pre-factor can differ by up to an order of magnitude depending on the 
approximations used in calculating the torque \citep{b76}, however for most disk models eqn.~(\ref{eq:qcrit}) is sufficient to estimate the minimum gap-opening mass 
\citep{edgar07}. Clearly, only IMBH or larger mass BH can open gaps in AGN disks. In particular, to open a gap requires low disk viscosity such 
that \citep{b76}
\begin{equation}
\alpha < 0.09 q^{2} \left(\frac{H}{r}\right)^{-5}.
\label{eq:opengap}
\end{equation}
Even if an IMBH is sufficiently massive to open a gap in a disk ($q>10^{-4}$ 
typically), the gap can close by pressure if the disk is geometrically thick enough such that \citep{bry}
\begin{equation} 
\frac{H}{r} >\frac{1}{40}\left(\frac{q}{\alpha}\right)^{1/2}.
\label{eq:pressure}
\end{equation}
Combining these conditions allows gap opening to occur when $H/r \lesssim$min[$(1/40)(q/\alpha)^{1/2},(q^2/\alpha)^{1/5}$]. Thus, any gap detection can tell us a lot about AGN disks. We do not address in detail the recent result from \citep{zhu13} that low-mass satellites may open gaps in a disk but note that in hot AGN disks, where MRI is effective, even IMBH with $q < 10^{-4}$ may open gaps.

 Just before gap-opening occurs, the disk perturbation may allow for very rapid Type III migration \citep{b28,b25,b26} with consequences for observed AGN luminosity \citep{b96}. Once a gap opens, the outer edge of the gap is exposed to the central radiation source and both the inner edge and the disk immediately outside the outer edge are shadowed, leading to a change in the observed spectral 
energy distribution \citep[e.g.][]{b60,b59} (and see \S\ref{sec:gap} below). The gap will be approximately 2$R_{H}$ in width where $R_{H}=a(q/3)^{1/3}$ is the IMBH Hill radius and $a$ the IMBH semi-major axis. The gap is not completely empty as the IMBH is fed by leading and trailing resonance arms (analagous to gap-opening Jupiters in protoplanetary disks) and the IMBH may have an accretion disk of size-scale $R \ll R_{H}$ \citep{hayasaki08,roedig12,dhm13}.  A gap-opening IMBH migrates within the disk on the so-called Type II migration timescale given by
\begin{equation}
\tau_{\rm II} =\frac{1}{\alpha} \left(\frac{H}{r}\right)^{-2} \frac{1}{\omega}
\label{eq:t2}
\end{equation}
where $\omega$ is the Keplerian angular frequency and $\tau_{\rm II}=\tau_{\alpha}$, the 
viscous disk timescale. 

The coevolution of an AGN disk with an IMBH can differ from that of planets in a protoplanetary disk. First, the Type II migration rate varies depending on the ratio of the gas to radiation pressure in the inner AGN disk, since this influences (H/r) and $\alpha$. Second, the mass of the IMBH may be larger than the local disk mass in the radiation pressure dominated regime. In this case, the disk banks up on the outer edge of the gap before it can push the IMBH inwards \citep{b91,ipp99,khl12}. The viscous timescale is smallest inside the orbit of the IMBH, so the inner disk can drain away, leaving an empty central cavity \citep{al96,b24}. The viscous disk will pile-up just outside the IMBH orbit, analagous to the build-up of water behind a dam. The dam may leak or burst after a time, refilling the inner disk and leaving an annulus in the disk and/or an accreting IMBH \citep[e.g.][\& references therein]{khl12,mck13a,dhm13,farris13}. Third, GW emission by the IMBH eventually dominates over Type II migration. If there is a circular cavity in the disk, once the GW timescale becomes shorter than
 the viscous time, the outer edge of the gap cannot follow the IMBH and effectively freezes until the binary merges \citep{an02,lwc03,mp05}. However, if there is an inner disk as the IMBH inspirals due to GW emission, the gap may retain an annular geometry with characteristic width $\sim R_{\rm H}$ tracking the inspiralling IMBH (\citealt{brm12}, see however \citealt{c10}). Only IMBH or SMBH can open gaps in AGN disks, so the detection of gaps in disks will set excellent constraints on $(H/r),\alpha$ in models of AGN disks \citep{ipp99,hayasaki08,rafikov12,khl12}, in spite of uncertainties in those models. 

\section{Gaps \& cavities in outer AGN disks: predictions}
\label{sec:gap}
Our model of IMBH growth displays strong parallels with models of giant planet 
growth in protoplanetary disks. A large fraction ($>1/5$) of protoplanetary disks exhibit evidence for gaps or cavities probably carved out by giant protoplanets \citep[e.g.][and references therein]{b88}. If AGN disk conditions permit an IMBH to carve out a gap or cavity, the AGN should display features and variability analagous to those in gapped protoplanetary disks. 

\subsection{SED dip: predictions}
\label{sec:disk_gap}
Consider an AGN disk consisting of annuli, each at a different temperature. In an AGN disk without a gap, each of the annuli contributes a black body spectrum to the overall multi-colour optical/UV spectrum. If the AGN disk includes a gap, then the black body 
spectrum due to that missing annulus is subtracted from the overall spectrum, leading
 to a dip or break around that disk temperature \citep{gult12}, independent of breaks due to reddening or absorption edges \citep[e.g.][]{zheng95} . From protoplanetary disk theory, the width of the gap ($w$) will be $ w \geq 2R_{H}=2(q/3)^{1/3}a \sim 0.07(0.14)a$, for $q=10^{-4}(10^{-3})$ \citep{b20}, where $a$ is the semi-major axis of the IMBH orbit. 

The spectral break wavelength ($\lambda_{b}$) due to a fully empty blackbody annulus in a homogeneous thin disk around a SMBH of mass $M$ accreting at rate $\dot{M}$ is 
\begin{equation}
\lambda_{b}=\left(\frac{hc}{xk}\right)\left( \frac{G}{2\pi\sigma}\right)^{-1/4}\left(\frac{M \dot{M}}{r^{3}}\right)^{-1/4}
\end{equation}
where $x \sim 5$ and we rewrite $\lambda_{b}$ as
\begin{equation}
\lambda_{b} \sim 140 \eta^{1/4} \left( \frac{r_{2}}{r_{g}}\right)^{3/4} \left( \frac{M_{1}}{10^{8}M_{\odot}}\right)^{1/4} \left( \frac{\dot{m}}{0.01}\right)^{-1/4} \rm{\AA}
\label{eq:lambreak}
\end{equation}
where $r_{2}$ is the location of the gap (in units of $r_{g}=GM_{1}/c^{2}$), $M_{1}$ is the mass of the primary SMBH, $\dot{m}$ is the Eddington accretion ratio ($\dot{m}=1.0$ is the Eddington rate) and $\eta$ is the accretion efficiency ($\eta=0.06-0.42$ for the full range of black hole spins). Thus, a gap at $10^{3}r_{g}$ in a thin homogeneous disk around a $10^{6}(10^{8})M_{\odot}$ SMBH leads to a break at $\lambda_{b} \sim 0.4(1.4)(\eta/0.1)^{1/4}\mu$m. Eqn.~\ref{eq:lambreak} can be compared directly with eqn.(13) in \citep{gult12}, with our pre-factor agreeing with theirs ($\sim 140$) for $\eta \sim 0.1$ if their $f(w/h) \sim 0.6$. A sufficiently wide gap leads to a broad dip in the broadband optical continuum (see e.g. Fig.~1 in \citet{gult12} for an illustration,  although their break should be located around $\sim 2\mu$m not $\sim 0.2\mu$m). Local disk mass is expected to decrease rapidly at small radii, so we should observe deeper and wider spectral dips in the SED as $\lambda_{b}$ decreases. 

Assuming the irradiated outer disk flares, such that $H/r$ is an increasing function of radius, the blackbody temperature of the irradiated outer gap wall ($T_{\rm wall}$)  is given by \citep{b20} 
\begin{equation}
T_{\rm wall}=L_{\rm inner}^{1/4}\theta^{1/4}r_{\rm wall}^{-1/2}
\label{eq:tcav}
\end{equation}
where $L_{\rm inner}$ is the AGN luminosity due to material at $r<r_{\rm wall}$ and $\theta=-H_{r}/r +dH_{r}/dr$ is the angle between the continuum source and the tangent to the disk surface and $H_{r} \sim H$ the disk thickness in the limit of a very optically thick disk. Rewriting in terms of $\dot{m},M_{1}$ we find the peak wavelength of outer gap emission is
\begin{equation}
\lambda_{\rm wall} \approx 165 \left( \frac{r_{2}}{r_{g}}\right)^{1/2} \left( \frac{M_{1}}{10^{8}M_{\odot}}\right)^{1/4} \left( \frac{\dot{m}}{0.01}\right)^{-1/4} \theta^{-1/4} \rm{\AA}
\label{eq:peak}
\end{equation}
where $\theta$ may be rewritten in terms of the opening angle of the disk at the wall ($\beta$) and radius of the inner accretion disk $r_{\rm in}$ as $\theta^{-1/4}\approx (cos \beta/4)^{1/4} (r_{\rm in}/r_{\rm wall})^{1/2}$. In a (more realistic) flared disk, $\lambda_{b}$ due to the gap is found from the $r^{-1/2}$ dependence in eqn.~\ref{eq:peak} rather than the $r^{-3/4}$ dependence in eqn.~\ref{eq:lambreak}.  Thus, a gap at $10^{3}\pm 10^{2}r_{g}$ in a flared disk around a $10^{6}(10^{8})M_{\odot}$ SMBH yields a break centered on $\lambda_{b} \sim 0.06(0.22)(\theta/30)^{-1/4}\mu$m and a corresponding bump due to the wall at $\lambda_{\rm wall} \sim 1.05 \times \lambda_{b}$.

In the gas pressure dominated region of the disk, the outer wall height can be approximated by \citep{b20} 
\begin{equation}
H_{\rm wall}  \approx \frac{c_{s}}{\omega} =\left(\frac{k_{\rm B}}{\mu m_{p}}\right)^{1/2}T_{\rm wall}^{1/2}\omega^{-1}
\end{equation} 
where $c_{s}$ is the sound speed \citep[see][for the radiation pressure dominated case]{b24}. The maximum luminosity of the outer wall can be approximated by $L_{\rm wall,max} \approx (H_{\rm wall}/r)  L_{\rm inner}/4$. Shadowing by the inner disk will reduce the observed value of $L_{\rm wall}$. For $H_{\rm wall}/r \sim 0.1$, then $L_{\rm wall} \leq 3\%L_{\rm inner}$. 

In certain cases a cavity will form in the AGN disk rather than a gap. When the local disk mass $<M_{2}$, secondary migration will stall \citep[e.g.][]{b91,khl12,mck13a}. As the inner disk drains 'inside-out', gas will pile-up at the outer gap edge and a cavity can form \citep{mck13a}. As the inner disk drains to form a cavity, $L_{\rm inner}$  decreases. Since our model predicts that in flared disks with gaps/cavities, 
\begin{equation}
\lambda_{\rm wall} \approx 140 \left(\frac{L_{\rm inner}}{10^{43}\rm{erg/s}}\right)^{-1/4}\left(\frac{\theta}{30}\right)^{-1/4}\left(\frac{r}{r_{g}}\right)^{1/2}\left( \frac{M_{1}}{10^{8}M_{\odot}}\right)^{1/2}\rm{\AA} 
\end{equation}
for $L_{\rm inner} \sim 10^{42(43)}$ergs/s, $\lambda_{\rm wall} \sim 0.8(0.4)\mu$m at $10^{3}r_{g}$. Our model predicts that while $L_{\rm inner}$ decreases, the SED slope at $\lambda \geq \lambda_{\rm wall}$ steepens due to continued viscous gas inflow and pile-up as $L_{\rm wall}$ increases. Thus, in disks with gaps/cavities, our model predicts an anti-correlation between average $L_{\rm inner}$ and the SED slope at $\lambda \geq \lambda_{\rm wall}$. This prediction distinguishes our model from models of cavity formation due to inner disk instability and collapse, since there is no (or little) pile-up in the latter case. Interestingly, pile-up at the outer gap edge will generate a luminous annulus in the disk, which may be detectable in an optical stellar transit of the AGN disk (see \S\ref{sec:transits} below).

As the inner disk drains inside-out, regions of short timescale variability are removed from the AGN disk. Gas pile-up on the viscous timescale ($\tau_{\alpha}(r_{\rm wall})$) will be accompanied by disk drainage on much faster inner disk timescales. Thus, our model predicts that a decline in power on short timescales in the power density spectrum (PDS) of an AGN, due to removal of the inner disk, will be accompanied by a prominent peak in the PDS due to variation on the characteristic pile-up timescales. If we approximate the gas pile-up on the gap/cavity edge as an unstable, thick advective flow  (particularly as photoionization declines), the timescale of variability ($\Delta t_{\rm wall}$) of such a flow is roughly the freefall timescale
\begin{equation}
\Delta t_{\rm wall} \geq 16 \left(\frac{M_{1}}{10^{8}M_{\odot}}\right) \left(\frac{r_{\rm wall}}{100r_{g}}\right)^{3/2} \rm{days}
\label{eq:cav_var}
\end{equation}
which is comparable to observed prominent breaks on timescales of $\sim 5-100$ days in AGN PDSs \citep[e.g.][]{collpet01}. Even if the pile-up is unstable on short times and leaks completely into the gap/cavity, refilling the disk, torques from the secondary can still excavate a new gap/cavity in the disk and the cycle begins again.

The dam wall that holds back the piled-up gas may be leaky at best and not endure for long, particularly at high accretion rates \citep[see][for extensive discussions]{mck13a}. In the case of high accretion rates, cavities may be short lived and quickly refilled as stalled gap-opening migration resumes. Cavities will only persist until mass $\geq M_{2}$ builds up at the cavity edge, 
so the cavity lifetime is $\tau_{\rm cav} \geq M_{2}/\dot{M}_{1}$. The AGN disk lifetime is given by $\tau_{\rm disk}=q_{\rm disk} M_{1}/\dot{M}_{1}$, so among disks with IMBH, a fraction $\tau_{\rm cav}/\tau_{\rm disk}=q_{2}/q_{\rm disk}$ will exhibit cavities where $q_{2}=M_{2}/M_{1}$ and $q_{\rm disk}=M_{\rm disk}/M_{1}$. Thus, if an AGN disk ($ q_{\rm disk} \sim 10^{-2}$), hosts a sufficiently massive IMBH ($q_{2} \sim 10^{-4}$) it has about a $1\%$ chance of being observed with a cavity. 

Disks around lower mass SMBH are the likeliest observational targets for finding gaps or cavities. Around low mass SMBH, note that even large mass (short-lived) stars could maintain a cavity in the inner disk for most of their main sequence lives.
 Around an SMBH with $M_{1}=10^{6}M_{\odot}$, an AGN disk lasts  $\tau_{\rm disk} \sim 50(q_{\rm disk}/10^{-2})(\eta/0.1)(M_{6}/10^{6}M_{\odot})(\dot{m_{1}}/0.01)^{-1}$Myr, where $\dot{m}_{1}$ is the Eddington ratio of the primary. A $M_{2}=10M_{\odot}$  stellar mass BH growing in this disk at $\dot{m}_{2}\sim 3.5\times$Eddington via collisions and gas accretion reaches gap/cavity opening threshold ($q_{2} \sim 10^{-4},M_{2}\sim 100M_{\odot}$ in time $\tau_{\rm gap} \sim 35(\dot{m}_{2}/3.5)$Myr \citep{b93}. Assuming every low mass AGN disk starts off with a stellar mass black hole, if $\tau_{\rm gap}< \tau_{\rm disk} $, a crude estimate of the probability of a gap or cavity in disks around $M_{1}\sim 10^{6}M_{\odot}$ SMBH is $P_{\rm gap}=1-(\tau_{\rm gap}/\tau_{\rm disk})$ or
\begin{equation}
P_{\rm gap}=1-\left[0.7\left(\frac{\dot{m}_{2}}{3.5}\right)\left(\frac {10^{-2}}{q_{\rm disk}}\right)\left(\frac{0.1}{\eta}\right)\left(\frac{10^{6}M_{\odot}}{M_{6}}\right)\left(\frac{\dot{m_{1}}}{0.01}\right)\right].
\label{eq:pgap}
\end{equation} 
A more detailed estimate can be constructed by accounting for the location of $M_{2}$, the local mass supply and Bondi accretion \citep{kyl11,ju+13}.
Thus, around low luminosity Seyfert AGN with $M_{1} \sim 10^{6}M_{\odot}$, we expect $\sim 1/3$ of disks to display evidence of gaps and only $\sim 1\%$ to display evidence for cavities. If IMBH in AGN disks start from stellar mass black hole seeds then only $\sim 1\%$ of higher luminosity Seyfert AGN will display evidence for gaps and little or no evidence for cavities due to IMBH formed in the disk. However, if IMBH survive the AGN phase, they can grow much larger in subsequent phases and a larger fraction of higher luminosity and higher mass AGN may display evidence for gaps and cavities. At $z<0.1$, there are tens of well-studied X-ray selected Seyfert AGN including $\sim 10$ low mass Seyferts. Several of these AGN (e.g. MCG-6-30-15) are heavily absorbed in the optical/UV, so it is difficult to constrain optical accretion disks in these cases. A systematic survey of hundreds of optically selected AGN (e.g. from SDSS) will yield the strongest statistical constraints on the occurrence of gaps and cavities in AGN disks. If, after accounting for reddening and naturally occurring spectral breaks \citep[e.g.][]{zheng95},  there is an absence of breaks in the 'big blue bump' of AGN SEDs, we can rule out the presence of close MBH mergers in AGN and severely restrict models of IMBH in AGN disks.

\subsection{Predictions from a parallel with gapped protoplanetary disks}
\label{sec:trans}
The previous section is based on physical expectations but we can also make predictions by analogy, based on observations of gapped protoplanetary disks. First, some emission lines in protoplanetary disks are observed to be double-peaked due to the presence of a gap or cavity, which leads to blue- and red-shifted peaks in the line emission \citep{b88}. By analogy, our model predicts double-peaked low ionization optical/UV line profiles in AGN. A small fraction of AGN are indeed observed to display double-peaked low-ionization emission lines \citep{b51,b52}. These lines and their variability \citep{b52} can be accounted for by a gap-opening IMBH trailed and led by clumpy spiral density waves. Double-peaked lines observed in NGC~4151 for example, could be explained by a black hole binary, with a large mass secondary ($q\sim 0.01-0.1$) at large eccentricity \citep{bon12}. Our model predicts that double-peaked lines must originate outside the disk cavity, so a spectral break or dip in the SED must occur at $\lambda_{b}<\lambda_{L}$, the line wavelength. We can test this prediction in the sample of AGN with double-peaked lines, by comparing the timescales of variability of the double-peaked lines ($t_{\rm l}$) with the timescale of variability of the outer gap wall ($t_{\rm wall}$). If $t_{\rm l} \leq t_{\rm wall}$ then our interpretation of these lines must be incorrect. 

Second, in protoplanetary disks, the outer gap/cavity wall is directly 
exposed to the stellar ionizing continuum and emits a 
blackbody at $\lambda_{\rm wall}$ with luminosity proportional to the wall height \citep{b88}, followed by a luminosity drop at $L(\lambda> \lambda_{\rm wall})$ due to shadowing by the wall. Our model predicts the same basic observable: a prominent blackbody peak in the AGN SED at $\lambda_{\rm wall}$,  with $L(\lambda>\lambda_{\rm wall})<L(\lambda_{\rm wall})$ due to shadowing. The luminosity at $\lambda_{\rm wall}$ varies on timescale $\Delta t_{\rm wall}$ which will be significantly faster than the timescale of variability at $\lambda>\lambda_{\rm wall}$.

Third, in the SEDs of gapped protoplanetary disks, as L($\lambda <\lambda_{\rm wall}$) increases, $L(\lambda>\lambda_{\rm wall})$ is observed to decrease, and vice versa \citep{b59}. This 'see-saw' variability is believed to be due to occasional puffing up of the inner disk (short wavelengths), shadowing the outer disk wall (long wavelengths). By analogy, our model predicts exactly this sort of spectral variability in gapped AGN disks. We predict that during a flaring high state in X-rays from the innermost disk, $L(\lambda < \lambda_{\rm wall})$ due to the irradiated inner disk must increase. Since the puffed up inner disk shadows the outer gap, $L(\lambda_{\rm wall})$ must decrease. Conversely, during an AGN 'low state' where $L_{x}$ diminishes, $L(\lambda < \lambda_{\rm wall})$ will also decrease, but $L(\lambda_{\rm wall})$ must increase as the outer gap/cavity wall is no longer shadowed by the puffed-up inner disk. Thus, in sources that are X-ray luminous (i.e. with a substantial inner disk), but with high flaring or low states, our model predicts a linear correlation between $L_{x}$ and $L(\lambda <\lambda_{\rm wall})$, but an anti-correlation between $L_{x}$ and $L(\lambda_{\rm wall})$.

\subsection{LINERs as AGN disks with cavities}
\label{sec:liners}
A sub-class of galactic nuclei, LINERs, do not exhibit a prominent blue/UV bump in their SEDs \citep[e.g.][]{bmaoz,b40}. Thus either LINERs do not have a prominent thermal disk or the accretion flow is radiatively inefficient. If there is no accretion disk, photoionization could be powered by AGB stars \citep[e.g.][]{erac10}. We suggest an alternative possibility. In our model, a substantial cavity carved out by an IMBH (or SMBH), can account for the absence of a prominent disk signature. By removing  the inner disk to large radii, the reduced ionizing continuum luminosity will change the optical line ratio from AGN-like. If the massive binary in the cavity is accreting weakly due to a leaky dam (see Sec.~\ref{sec:observations}), the ionizing continuum will have a much lower luminosity, akin to LLAGN or an ultra-luminous X-ray source (ULX) and therefore the observed line ratios must be LINER-like \citep{b99}. ULXs (which may be powered by accretion onto IMBH) have now been observed with optical line ratios remarkably consistent with those in LINERs and LLAGN \citep{b61}. Thus, our model predicts that many LINERs consist of MBH binaries in large central cavities in a gas disk. A small fraction of these LINERs will be GW loud (depending on $a_{\rm bin}$).

Specifically our model predicts: (1) a rise in the optical spectrum of LINERs due to the cavity wall at $\lambda_{\rm wall}$, together with (2) an immediate dip at $\lambda>\lambda_{\rm wall}$ due to pile-up and disk shadowing of a weakly accreting  central source, (3) significantly shorter variability timescales at  $L(\lambda_{\rm wall})$ compared to $L(\lambda > \lambda_{\rm wall})$, (4) little or no power in the LINER PDS on timescales $<t_{\rm wall}$ and (5) double-peaked low ionization lines at $\lambda > \lambda_{\rm wall}$. In this model, the IMBH/SMBH occupy a substantial cavity. If the cavity is not leaky the binary will not accrete strongly. Such LINERs will not exhibit a prominent, luminous, soft X-ray excess (see \S\ref{sec:disk_nogap} below). However, if the cavity is leaky, accretion onto the IMBH/secondary MBH could dominate the accretion onto the SMBH yielding ULX-like optical line ratios. 

\section{Gaps \& cavities in inner AGN disks: predictions}
\label{sec:feka}
\begin{figure}
\begin{center}
  \resizebox{\columnwidth}{!}{\includegraphics{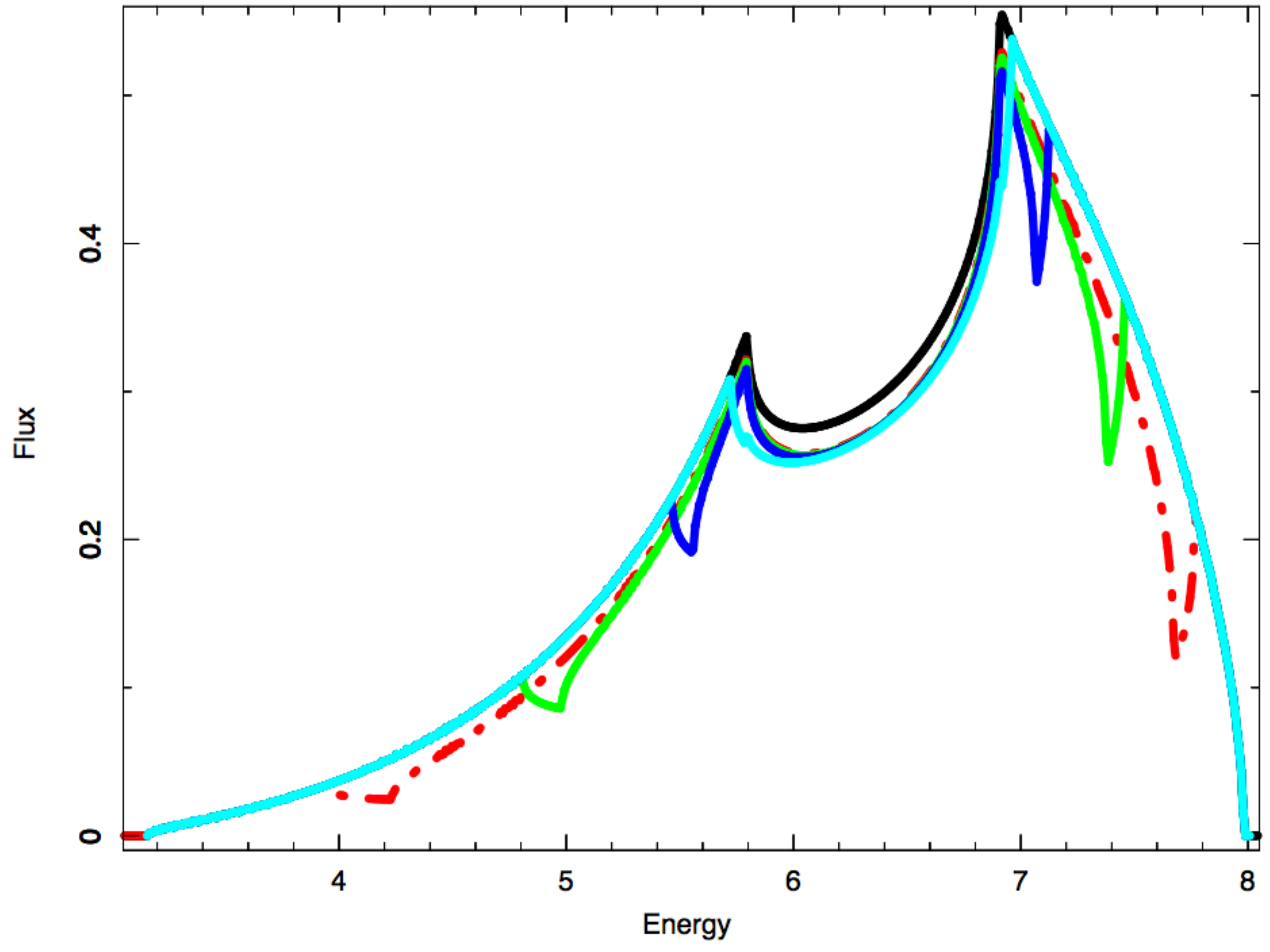}}
\end{center}
\caption[]{The change in Fe~K$\alpha$ profile for the in-migration of a $q=3 \times 
10^{-3}$ IMBH across the inner disk with empty gap of width $2R_{H}$ where 
$R_{H}=(q/3)^{1/3}R$. Flux is in normalized arbitrary units and Energy is in units of keV. 
Solid black line indicates the unperturbed Fe~K$\alpha$ 
profile. Turquoise solid line denotes an annulus at $90\pm9r_{g}$ in the inner 
disk. Dark blue solid line denotes an annulus at $50\pm 5 r_{g}$, Green solid line 
corresponds to an annulus at $20\pm 2r_{g}$ and the dashed red line corresponds to 
an annulus at $10\pm 1r_{g}$. See \citet{mck13a} for detailed discussion of this and other effects in the broad FeK$\alpha$ line as well as detectability with future missions.
\label{fig:feka}}
\end{figure}

In the limit of small disk radii, as a gap-opening IMBH migrates closer to the supermassive black hole, the gap in the accretion disk can imprint itself on the broad component of the Fe~K$\alpha$ line profile \citep[see][for extensive discussion]{mck13a}. Fig.~\ref{fig:feka} illustrates the simulated, noiseless variation in a broad Fe~K$\alpha$ line profile due to an in-migrating empty annulus (occupied by an IMBH) of width $2R_{H}$ (where $R_{H}=(q/3)^{1/3}a)$ and a is the circularized orbit radius) decreasing from $100r_{g}$ to $6r_{g}$. We calculated the broad 
Fe~K$\alpha$ line profile around a Schwarzschild black hole using the \texttt{diskline} algorithm \citet{fab}. We assumed $q=3 \times 10^{-3}$, the disk is inclined by $60^{o}$ to the observer's line of sight and the X-ray continuum goes as $r^{-2.5}$. As 
the IMBH and gap migrate into the inner $100r_{g}$ of the disk, a flux deficit appears (turquoise solid line) compared to the unperturbed profile (black solid line). As 
inward migration continues, the empty gap removes flux from increasingly red and blue regions of the disk and the twin notches move apart, red-ward and blue-ward respectively from the line center (6.4keV). As the empty gap moves inwards in the disk 
from 50$r_{g}$ to 10$r_{g}$ (red curve), the 
blue notch moves from 6.5keV out to 7.2keV, and the red notch moves 
from $5.6$keV to 4.2keV. 

The timescale to merger of an IMBH-SMBH binary via gravitational radiation is given by \citep{b49}
\begin{equation}
\tau_{\rm GW} \approx 10^{12} \rm{yr} \left(\frac{10^{3}M_{\odot}}{M_{2}}\right)\left(\frac{10^{6}M_{\odot}}{M_{1}}\right)^{2} \left(\frac{a_{\rm bin}}{0.001 \rm{pc}}\right)^{4} (1-e^{2})^{7/2}
\label{eq:gravrad}
\end{equation}
where $M_{2}$ is the mass of the secondary IMBH, $M_{1}$ is the mass of the primary supermassive black hole, $a_{\rm bin}$ is the binary separation and $e$ is the IMBH orbital eccentricity. For a $3000M_{\odot}$ IMBH around a 
$10^{6}M_{\odot}$ supermassive black hole, the progression from green to red curve in 
Fig.~\ref{fig:feka} takes $\sim 16$yrs and from red curve to final merger takes $\sim 1$yr. Low 
mass AGN exhibiting broad Fe~K$\alpha$ lines (e.g. MCG-6-30-15) have the shortest timescales for merger and low mass migrators (IMBH and stellar mass BH) in the inner disk are the most likely sources of a late-stage ripple effect \citep{mck13a}. Oscillations caused by e.g. an accreting secondary (IMBH) in a cavity around a primary SMBH can be readily tested with the proposed \emph{LOFT} mission (see \citet{mck13a} for further details). Future X-ray missions such as \emph{Astro-H} and \emph{IXO-Athena} have the energy resolution to distinguish between features  due to highly ionized Fe \citep[e.g][]{b90} and may detect the ripple signatures or oscillations.  Observations of the latter stages of the ripple effect in the broad FeK$\alpha$ line of a low mass AGN, predict a prompt rise in gravitational wave (GW) luminosity from this source, which should reach the threshold of future GW detectors. This signal therefore constitutes a prime EM precursor for GW observations.

\section{Embedded objects in AGN disks: predictions}
\label{sec:disk_nogap}
In this section, we discuss the case of embedded objects in AGN disks, including IMBH that do not open a gap. In \S\ref{sec:softxs} we discuss the observational signatures of accreting embedded objects in the AGN disk. In \S\ref{sec:retro} we discuss observational signatures due to embedded objects on retrograde orbits within the AGN disk. In \S\ref{sec:xsbat} we examine whether one or other or both models could contribute to, or correspond wholly to the soft X-ray excess observed in many AGN. 

\subsection{Predictions from accretion}
\label{sec:softxs}
IMBH can accrete from the gas disk at up to Eddington rates \citep{b35}, although they can grow at super-Eddington rates including collisions in the disk \citep{mck13a}. A gas disk accreting onto an IMBH produces a  blackbody spectrum peaking in the soft X-ray band at temperatures of $\sim 0.1 \rm{keV}(10^{4}M_{\odot})-1$keV($10^{2}M_{\odot}$). This is observationally 
interesting because in many (possibly most) AGN, an 
excess of soft X-rays ($\sim 0.1-1.0$keV) is observed, relative to the level 
expected from extrapolating a powerlaw fit from hard X-rays ($>1$keV) 
\citep[e.g.][]{b55,scott12}. The excess is observed to have a 
quite remarkable constant temperature peak, independent of large variations in 
observed AGN luminosity \citep{b67,b6}. The 
origins of the soft excess remain unknown, however accretion onto an IMBH (or 
population of IMBH seeds) in the AGN disks could account for the luminosity and 
energy of the soft excess \citep{b98,b99}. 

From \S\ref{sec:gap}, the innermost disk blackbody temperature for a thin disk around an IMBH of mass $M_{2}$, is $T \propto M_{2}^{-1/4}$, 
with luminosity $L \sim \eta \dot{M}_{2} c^{2}$, where $\eta$ is the accretion 
efficiency. For black holes ($10-10^{4}M_{\odot}$), $T \sim 10^{7}-10^{6}$K 
(or $\sim 1-0.1$keV) and $L \sim \eta 10^{39}-10^{42} \dot{m}_{\rm Edd}$ erg$\rm{s}^{-1}$, where $\dot{m}_{\rm Edd}$ is the Eddington ratio. The soft X-ray excess in AGN peaks at $\sim 0.1$keV with $L_{sx} \sim 10^{42-43}$ erg $\rm{s}^{-1}$ \citep[e.g.][]{b9,b6}. Thus, accretion onto IMBH with masses 
$\geq 10^{4}M_{\odot}$ could account for the observed soft excess in AGN. In this model $L_{sx}$ increases as a function of $M_{2},\dot{M}_{2}$, since from eqn.~\ref{eq:lambreak} $T_{sx} \propto M_{2}^{1/4}\dot{M}_{2}^{1/4}$. Once an IMBH grows massive enough to open a gap ($q_{2} \geq 10^{-4}$, depending on $H/r,\alpha$),
we expect $L_{sx}$ to drop dramatically, unless the IMBH has an eccentric orbit \citep{roedig12}. In this model, AGN without a significant soft excess harbour  gap-opening IMBH or very low mass IMBH. 

The building blocks of IMBH in our model (nuclear cluster objects) will also accrete and generate (potentially) observable signatures. Some $10^{4}$ white dwarfs accreting at Eddington could produce a soft X-ray spectral bump $\sim 0.03-0.05$keV \citep{b38}  with luminosities $\sim 10^{42}$ erg $\rm{s}^{-1}$. Stars embedded in the AGN disk will look similar to T-Tauri stars, emitting X-rays at $\sim 10^{7}$K ($\sim 0.9$keV) but $L_{x}$ is only $ \sim 10^{28}-10^{32} \rm{erg}\rm{s^{-1}}$  \citep{b68}. So, even a large population ($\sim 10^{5}$) of T-Tauri stars in the disk would be insufficient to reproduce the AGN soft X-ray excess luminosity and a large fraction of the T-Tauri  luminosity will be reprocessed as IR.

\subsection{Predictions for objects on retrograde orbits}
\label{sec:retro}
Unlike protoplanetary disks, nuclear cluster objects (NCOs, i.e. stellar mass black holes, stars, stellar remnants) in AGN disks can follow retrograde orbits. Migration onto the SMBH will only occur if the retrograde NCO captures 'negative' angular momentum within its influence radius from a comparable mass of gas \citep[e.g.][]{nix11}. If most gas flows past the NCO, very little negative angular momentum will be transferred. Large-mass retrograde NCOs embedded in a disk (not in a cavity as in \cite{nix11}) will not open gaps because of lack of retrograde resonant torques. However, they may open a narrow annular gap as wide as their Bondi radius.  Retrograde NCOs and IMBHs could persist a long time at the same AGN disk radius due to far weaker torques than predicted by standard Type I and Type II migration scenarios (calculated for prograde orbiters). The observational signature of retrograde objects will be the 
X-ray generating bow shock associated with the headwind of NCO retrograde motion.

\begin{figure}
\begin{center}
  \resizebox{\columnwidth}{!}{\includegraphics{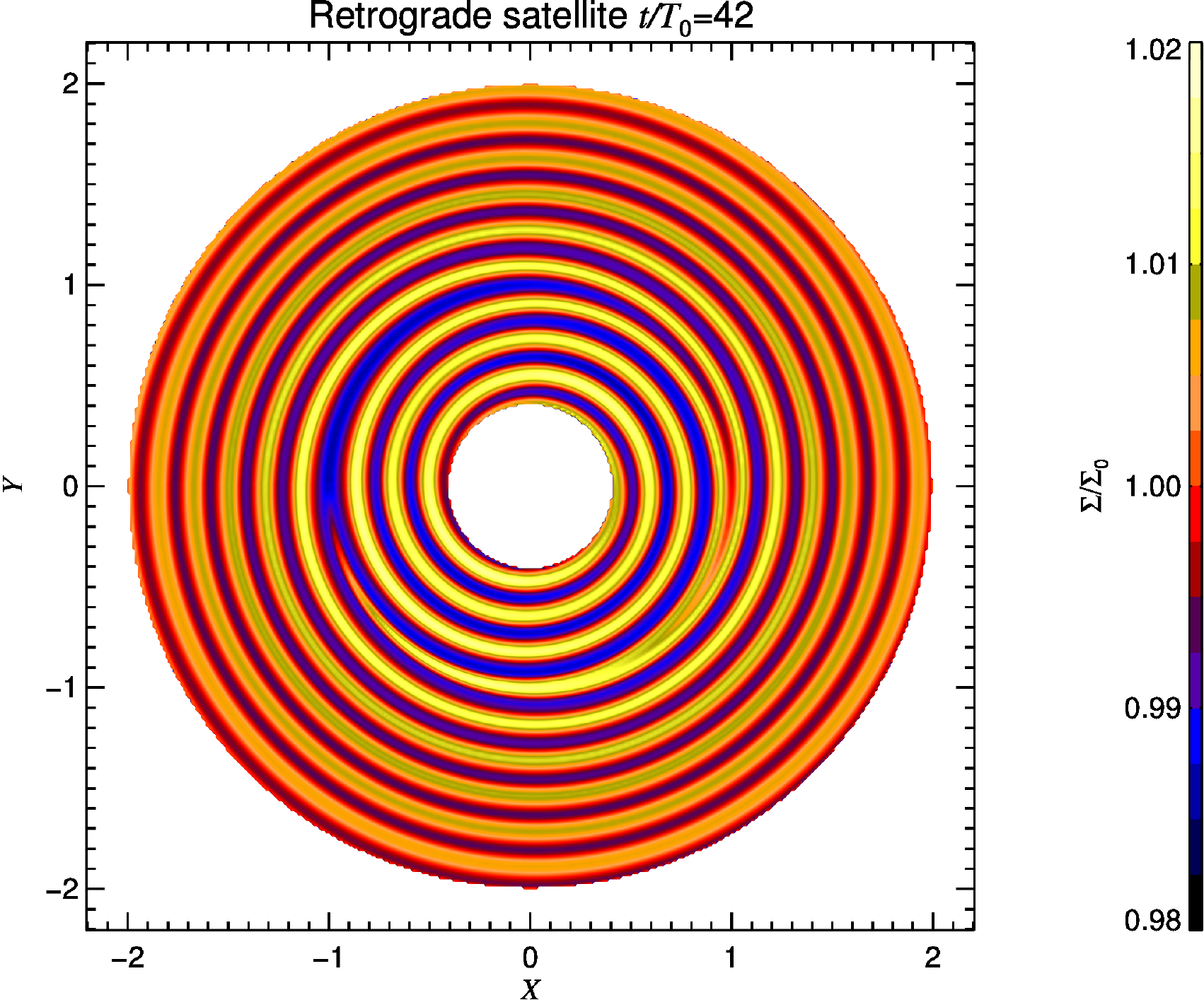}}
\end{center}
\caption[]{A 2-d simulation of an NCO with mass ratio $q=10^{-4}$ on a retrograde orbit in an isothermal, constant temperature disk of aspect ratio $H=0.05$ in a box of dimensions $r=[0.4,2.0],\phi=[-\pi,\pi]$. An identical prograde orbiter in the same disk, after the same time, opens a prominent gap and generates spiral density waves an order of magnitude larger than those here, leading to standard Type II migration.
\label{fig:retro_cart}}
\end{figure}

Fig.~\ref{fig:retro_cart} shows the results of a 2-d simulation using the {\sc Pencil Code} {\footnote{The code is publicly available under a GNU open source license and can be downloaded at http://pencil-code.googlecode.com}} of a satellite with  mass ratio $q=10^{-4}$ on a retrograde orbit in an isothermal, constant temperature disk of aspect ratio $H=0.05$ in a box of dimensions $r=[0.4,2.0],\phi=[-\pi,\pi]$. From Fig.~\ref{fig:retro_cart} the tapering spiral density waves we would expect to see in the prograde case are replaced with equal (but very small) magnitude density waves. Since the material in the tail increases by $\leq 2\%$ only over background after 42 orbits, the net torque of the gas on the retrograde NCO is far smaller than in standard (Type I or II)  migration scenarios, so the NCO remains at approximately the same disk radius. A bow shock caused by a star passing through a disk at relative velocity $v_{r}$ will heat gas behind the shock to $T\sim 0.1m_{p}v_{r}^{2}/k $ \citep{zent83}, where $m_{p}$ is the proton mass and $k$ is Boltzmann's constant. To generate bow shocks with temperatures $ \sim 0.1-1$keV, requires relative velocities in the range $v_{s} \sim 300-1000$km$\rm{s}^{-1}$, implying retrograde NCOs on Keplerian orbits in the outermost disk or torus ($\sim 10^{4-5}r_{g}$). The associated shock luminosity $L \approx \sigma A T^{4} \sim 10^{42-43}$ erg/s requires the area of $\sim 20$ Sun-like stars on retrograde orbits. Alternatively, the shock luminosity could be due to the single collision cross-section ($\sigma_{\rm coll} \approx \pi r_{g}^{2} (1+4GM_{2}/r_{g}v_{s})$ ) of a $M_{2} \sim 10^{3-4}M_{\odot}$ IMBH on a retrograde orbit in the outer disk. 
This model of the soft X-ray excess predicts little fractional variability in the soft excess as well as a constant soft excess luminosity, independent of changes in the continuum due to accretion onto the primary SMBH. This model can be ruled out if the luminosity and variability of the soft excess and the X-ray continuum are correlated, which can be tested for a large sample of nearby AGN with \emph{LOFT}.  

\subsection{Comparison with observations}
\label{sec:xsbat}
\begin{figure}
\begin{center}
  \resizebox{\columnwidth}{!}{\includegraphics{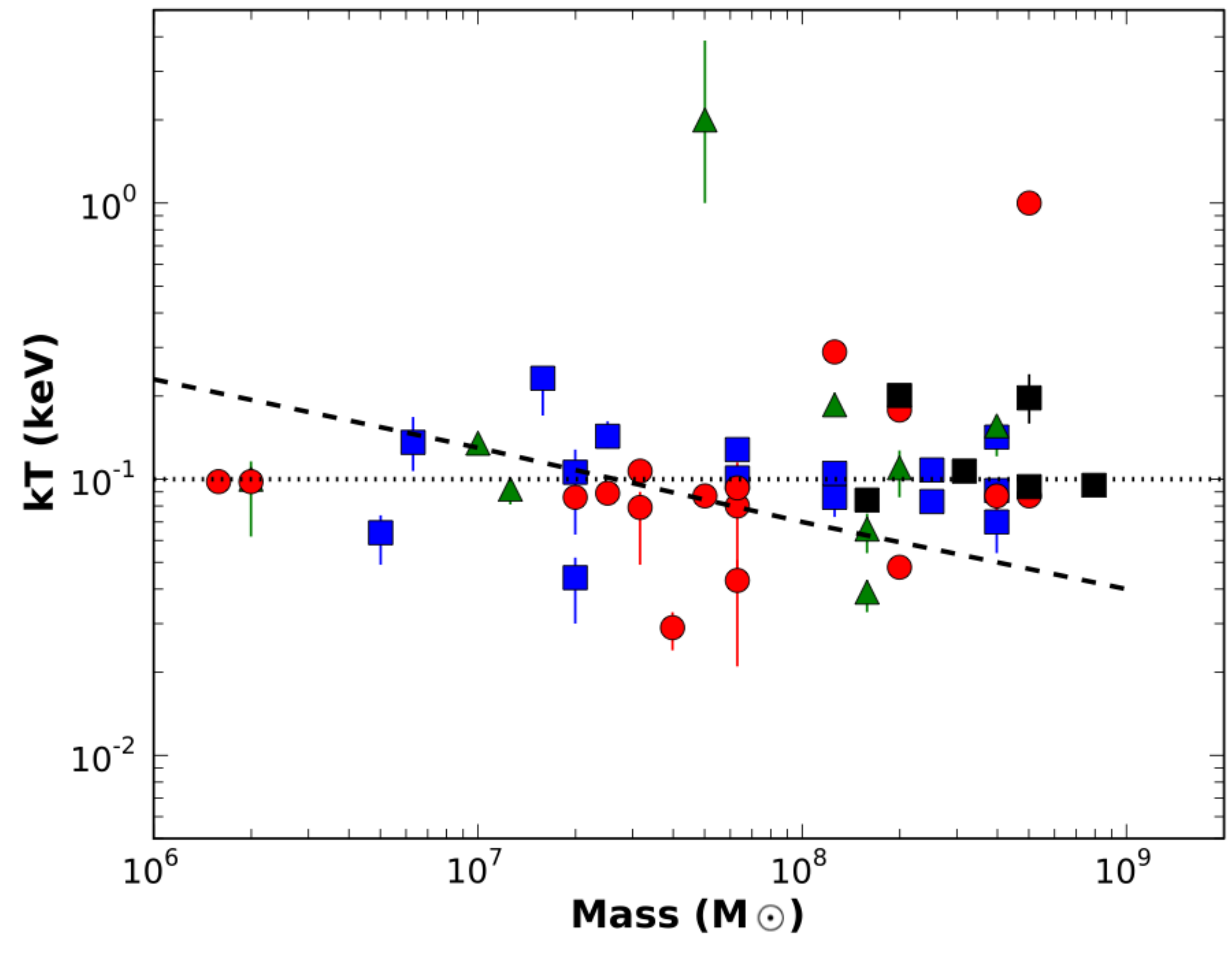}}
\end{center}
\caption[]{The peak energy (kT) of the soft X-ray excess found in AGN observed with 
Swift-BAT plotted against the mass of the central supermassive black hole (from \citet{b6}). Blue squares are Seyfert 1s, black squares are broad line radio galaxies, green triangles are Seyfert 1.2s and red circles are Seyfert 1.5s. The horizontal dotted line corresponds to a model of 
$\sim 0.1$keV thermal emission due to bow-shocks from NCOs/IMBH on retrograde orbits in the AGN disk, encountering 'headwinds' centered on $100$ km $\rm{s}^{-1}$.  The sloping dashed line corresponds to
IMBH with a constant mass ratio $q=2\times 10^{-4}$. Below this line, in AGN disks with fiducial $H/r=0.05,\alpha=0.01$ we expect no soft excess due to accreting IMBH (see text).
\label{fig:softxs}}
\end{figure}

The Swift Burst Alert Telescope (BAT) in the 14-195keV band is unbiased towards obscured sources and host galaxy properties \citep[e.g.][]{b6}. Fig.~\ref{fig:softxs} shows those AGN from the BAT sample that require a soft X-ray excess fit to the continuum \citep{b6}. As we can see, the result is a scatterplot, but centered around $\sim 0.1$keV. The two dashed-dotted lines in Fig.~\ref{fig:softxs} correspond to our two simple models of the soft excess. The horizontal dotted line corresponds to X-ray emission due to bow shocks centered on a constant velocity of $\sim 100$ km $\rm{s}^{-1}$ in the outer AGN disk, due to NCOs and IMBH on retrograde orbits. The sloping dashed line corresponds to the gap-opening threshold for an IMBH ($q \leq 2 \times 10^{-4}$, for fiducial disk parameters of $H/r \sim 0.05, \alpha \sim0.01$) as a function of primary SMBH mass. There is substantial scatter around both model fits, but the constant temperature model is a better global fit (confirmed by a simple $\chi^{2}$ test). The IMBH accretion model predicts weak or no emission from IMBH below the sloped dashed line because such objects should open a gap and decrease accretion. However, we cannot rule out this model based on Fig.~\ref{fig:softxs} alone since a small factor of $\sim 2-3$ on $H/r$ and $\alpha$, particularly in disks around lower mass SMBH, could account for soft excesses below the sloped dashed line. The retrograde orbits model can account for the range  of 43/45 of the AGN in Fig.~\ref{fig:softxs} if the  'median headwind' simply lies in the range $30-300$ km $\rm{s}^{-1}$ (implying most NCOs/IMBH are $>10^{3}r_{g}$ from the SMBH). In order to rule out one or both models, we need constraints from variability studies, using large effective area X-ray telescopes such as the future \emph{LOFT} mission.

\subsubsection{Test cases: Ton S180 \& future suggestions}
\label{sec:xsvar}
Constraining models of the soft excess is complicated by the presence of warm absorbing gas in AGN \citep{b97}. In order to distinguish between models of the soft excess, it helps to consider AGN with a soft excess but little or no warm absorption. For example, the narrow line Seyfert galaxy Ton S180 \citep{cv06} in a low state reveals a substantially stronger fractional variability in the softer X-rays than in hard X-rays \citep{b33}. In high states, the fractional variability of soft X-rays in Ton S180 appears to be similar to that of hard X-rays. In this AGN, we can rule out objects on retrograde orbits as a cause of the soft X-ray excess. If the soft excess is due to IMBH accretion, it evidently dominates emission during the low state while continuum emission dominates during the high state. However, rather than investigating individual idiosyncratic sources, it makes more sense to study a sample of AGN with soft excesses, but no warm absorption, particularly during low X-ray states where the continuum does not dominate. Observations of such a sample of AGN during 'low' states, with the large collecting area of \emph{XMM} or future missions such as \emph{LOFT}, will allow us to test the IMBH accretion model by comparing the variability timescales of the soft excess and the hard X-ray continuum. The accreting IMBH model predicts that the ratio of soft excess/continuum variability timescales during the low state is a function of the mass ratio (q) of the IMBH to the supermassive black hole. On one hand, competing models of the soft excess emission, such as blurred reflection \citep{b67} can be ruled out if it can be shown that the soft excess consistently varies on shorter timescales than the continuum.  On the other hand, we can rule out the presence of luminous, accreting IMBH in AGN disks if the fractional variability of the soft excess and the hard powerlaw components are identical for a large sample of these AGN in 'low' X-ray states.

\section{Intermittent IMBH spectral signatures:predictions}
\label{sec:indep}
Once-off (or intermittent) observational signatures of IMBH in AGN disks or (more likely) in post-AGN galactic nuclei will be observed infrequently, but can provide strong evidence for the presence of an IMBH in a galactic nucleus. From eqn.~\ref{eq:gravrad}, a $10^{2}M_{\odot}$ IMBH will remain orbiting a $10^{6}M_{\odot}$ SMBH for $\sim$Gyr if located $\geq 2000r_{g}$. IMBH may survive the AGN phase, in the same way that Jupiter-sized planets survive protoplanetary disks. Therefore we should expect to find IMBH in quiescent galactic nuclei (with enhanced stellar tidal disruption rates) and such IMBH may act as seeds for further growth in a new AGN phase.
 If IMBH are common in AGN disks (as we propose), surveys of hundreds of nearby galactic nuclei using e.g. LSST in the optical band will find the occasional event as described below.

\subsection{Transits of AGN disks:predictions}
\label{sec:transits}
An accreting IMBH in an AGN disk produces an asymmetric X-ray intensity distribution.
The asymmetry of the X-ray intensity distribution may be detectable by a transit of a 
bloated star or optically thick cloud across the face of an AGN \citep[see][for detailed transit calculations]{bb12}. The first such transit 
in an AGN was observed in MCG-6-30-15 \citep{b92}. This was followed by observations of transits in among others: NGC 3516 \citep{b7}, NGC 1365 \citep{b3} and Cen A \citep{b5}. A transit event, though rare, gives us a chance to map the AGN continuum at unprecedented resolution ($<$ micro-arcseconds). If the X-ray intensity profile is asymmetric, this will show up in the best-fit to the transit profile \citep{b92}. A systematic archival search of AGN X-ray lightcurves for transit profiles (a significant task far beyond the scope of this paper), will map the innermost accretion disk at unprecedented angular resolution and put strong limits on the occurrence of IMBH in AGN inner disks. In the optical band, transits can reveal rings of enhanced emission due to the pile-up of gas outside the orbit of a gap- or cavity-opening IMBH. 

\subsection{QPOs and spiral density waves}
\label{sec:imbh_qpo}
Quasi-periodic oscillations are occasionally observed in AGN and a fascinating possibility is that they arise due to the action of spiral density waves in the AGN disk \citep{b62}. Depending on the disk structure, migrating IMBH should have particularly strong associated spiral density waves, analagous to those associated with migrating giant planets in protoplanetary disks \citep{b20}. For example, a 3-4ks QPO is observed in RE J$1034+396$ enhancing the flux by $\sim 5-10\%$ in the 0.3-10keV X-ray band \citet{b62}. The orbital time at $\sim 6r_{g}$ around a $10^{7}M_{\odot}$ SMBH is $\sim 3$ks. For a $10^{5}M_{\odot}$ gas disk around a $M_{1}=10^{7}M_{\odot}$ SMBH we expect $\leq 1\%$ or $\sim 10^{3}M_{\odot}$ of gas within $10^{2}r_{g}$ from a standard thin disk model \citep{b21}. If this QPO is due to spiral arms from a migrating IMBH in the AGN disk, the IMBH is $M_{2} \leq 10^{3}M_{\odot}$ and $a_{\rm bin}\sim 10^{2}r_{g}$ from the SMBH. The IMBH will merge with the SMBH within $\sim 60(10^{3}M_{\odot}/M_{2})(10^{7}M_{\odot}/M_{1})^{2}(a_{\rm bin}/10^{2}r_{g})^{4}$kyrs. Since the co-rotating mass of gas at $r<10^{2}r_{g}$ is small, the IMBH will open a gap in the inner disk. RE J$1034+396$ is therefore a prime candidate for follow-up study of the broad component of the Fe~K$\alpha$ line (see \S\ref{sec:feka} above). Post-merger, we also expect a kick on the merged object to generate a density caustic that ripples outward in a spiral wave that decays with time, but only for mass ratios $q>10^{-2}$ \citep{b45}. A future timing mission like LOFT is needed in order to do a high cadence variability study of a large sample of nearby X-ray bright AGN. Spiral density waves raised by a migrating IMBH may show up as ripples in the broad FeK$\alpha$ line profile. Future timing studies of broad FeK$\alpha$ lines with the energy resolution and effective area of $\emph{Athena}$ will help constrain the location of IMBH migrators in AGN disks. 

\section{Gravitational waves: predictions}
\label{sec:gw}
Our model predicts a large number of merging nuclear cluster objects (NCOs) and IMBH in the accretion disk around a SMBH which may be rich sources of gravitational waves (GWs). A binary emits GWs with characteristic GW frequency \citep{pm63}
\begin{equation}
 f_{\rm GW} \sim \frac{(1+e)^{1/2}}{(1-e)^{3/2}} \sqrt{\frac{G M}{a^3}}\,.
\end{equation}
The detectable GW frequency band of the planned spaced-based GW observatory
LISA\footnote{\url{http://lisa.nasa.gov/}} or NGO\footnote{\url{http://elisa-ngo.org/}}
\citep{eLISA} is between $10^{-4}$ and $1\;$Hz,
and between $10$ to $3000\;$Hz for the existing Earth-based instruments
LIGO\footnote{\url{http://www.ligo.caltech.edu/}} and
VIRGO\footnote{\url{http://www.ego-gw.it/}}.
Binaries orbiting outside the last stable orbit, $a(1-e^2) \gtrsim (6+2e)\; r_g$,
are in the LISA (LIGO) frequency band if the total mass $M=m_1+m_2$ is between $10^3$--$10^7\;M_{\odot}$
($1$--$10^3\;M_{\odot}$). We discuss the GW signatures of various mass sources in turn below.

\subsection{GWs from an IMBH orbiting a SMBH}\label{s:GW1}
A compact object orbiting around a SMBH on a circular orbit emits GWs in the LISA band
if the orbital period is less than 6 hours. For SMBH mass $10^6\;M_{\odot}$
($10^7\;M_{\odot}$), this corresponds to an orbital radius less than
$75\;r_g$ ($16\;r_g$), or a time to merger less than
$100 \;\mathrm{yr}$ ($20\;\mathrm{yr}$) for an IMBH $\mu=10^3\;M_{\odot}$.
The signal to noise ratio (SNR) decreases with source distance 
and increases with $\mu$ as $S/N\propto \mu/D$. 
The detectable distace of a circular SMBH--IMBH binary with $S/N=10$ with LISA is very roughly\footnote{Here we assume that the LISA detector noise spectral amplitude decreases approximately
as $f^{-2}$ for $f\lesssim 10^{-3}$Hz \citep{bc04}, and used that
the dimensionless GW strain amplitude scales as $h\propto G^2 c^{-4} M\mu/(r D)$, 
according to the leading order quadrupolar radiation formula 
averaged over binary orientation. $D_\mathrm{LISA}$ varies by a factor 10 for $f\lesssim 10^{-2}\;\mathrm{Hz}$
due to the unresolved white dwarf background and different orientations.
}
\begin{equation}
 D_\mathrm{LISA} \sim 1\,\mathrm{Gpc} \left(\frac{M}{10^6M_{\odot}}\right)^{-2}
  \frac{\mu}{10^3M_{\odot}}
  \left(\frac{r}{30\,r_g}\right)^{-4}\left(\frac{T}{1\,\mathrm{yr}}\right)^{1/2}
\label{eq:DLISA}
\end{equation}
where $T$ is the observation time. Eqn.~\ref{eq:DLISA} implies we will detect all SMBH-IMBH mergers out to $z \sim 0.3$. The GW phase evolution may be used to independently detect and distinguish 
thousands of SMBH--IMBH binaries in the Universe with LISA, if they exist. 
The binary separation shrinks over timescale $\tau_\mathrm{GW}\propto r^4$
given by Eq.~(\ref{eq:gravrad}). The relative fraction of SMBH-IMBH binaries 
in the Universe, that reside in a logarithmic separation interval centered at $r$, 
is proportional to $\tau_\mathrm{GW}$ (Eqn.~(\ref{eq:gravrad})). Thus, most binaries will reside at 
large separations. However, the SNR increases toward smaller separations as $r^{-4}$, 
which implies that LISA will discover SMBH--IMBH binaries with a uniform probability 
distribution as a function of $\ln(r)$. However, at very small frequencies, the sources may constitute an unresolved GW background of $100M_{\odot}$ intermediate mass ratio inspirals \citep[see][for extreme mass ratio inspirals]{bc04b}.

Once LISA detects a SMBH--IMBH binary, it can measure its physical parameters 
from the GW signal including the masses, separation, eccentricity, sky location, 
and distance to the source to a level similar to SMBH-SMBH mergers. The sky localization precision is of order $1\;\mathrm{deg}^2$
for a SMBH-SMBH or SMBH-NCO binary approaching merger ($r \lesssim 10r_{g}$) at cosmological redshift $z=1$ \citep{bc04,lhc11,mkfv12}, and $10\;\mathrm{deg}$ or more 
at separations $r\gtrsim 20\,r_{g}$ \citep{lh08,khmf07}. If the SMBH-IMBH inspirals can be identified with a similar accuracy, the 3D source localization accuracy may be sufficient to identify a unique AGN or LINER counterpart to the GW source approaching merger \citep{kfhm06}. If not it will provide a $10\;\mathrm{deg}^2$ sky area for a deep triggered search for other electromagnetic signatures days to months before merger 
\citep[see][]{khmf07,khm08}. Bright AGN activity is not expected during merger if the torques of the secondary 
excavates a hollow cavity in the disk; instead we expect LINER-like activity (see \S\ref{sec:liners} above). However, depending on the 'leakiness' of the cavity wall, non-axisymmetric 
streams can supply gas to the inner regions, leading to a coincident 
electromagnetic counterpart (see discussion in \S\ref{sec:observations}). A precise measurement of the GW phase may also be used to look for perturbations 
caused by the astrophysical environment around the SMBH--IMBH binary. 
The LISA measurement accuracy is sufficient to detect the torques 
generated by the spiral density waves in the accretion disk if $\mu \lesssim 100 M_{\odot}$
\citep{kyl11,yklh11}.

\subsection{NCOs-IMBH \& IMBH-IMBH mergers}\label{s:GW2}
In our model the accretion disk is expected to host abundant stellar-mass NCOs. Upon close encounters with the IMBH, 
these objects may become bound to the IMBH on very eccentric orbits due to GW emission \citep{okl09}. Thus, our model \emph{predicts a new, unexpected source of GWs}. These systems generate repeated GW bursts, 
detectable with both LIGO and LISA coincidentally \citep{kl12}. This is possible 
if the orbital time is less than $6\;\mathrm{h}$ for LISA detections, 
and the pericenter passage timescale less than a $0.1\;\mathrm{s}$ for LIGO detections. As the eccentricity shrinks, the signal morphs into a continuous inspiral signal, and eventually a merger and ringdown. 
The detection range for LIGO and LISA is between $1$--$10\;\mathrm{Gpc}$ and $10\;\mathrm{Mpc}$, respectively \citep{east13}, so the odds of a coincident detection are remote. 

There are no studies on the parameter estimation accuracy of repeated burst sources to date. 
Arguably, it should be much better than for circular sources, 
since these sources are in the detectable frequency band for a much longer time, the 
signal power is enhanced at large frequencies due to eccentricity, and these sources 
exhibit apsidal precession which can break parameter degeneracies \citep{mkfv12}.
For circular sources IMBH--NCO binaries are outside the detectable range of existing
Earth-based observatories. Their detection requires the future third generation instrument, the Einstein Telescope\footnote{\url{http://www.et-gw.eu/}}. With the latter, circular IMBH--NCO binaries may be localized within 2 deg accuracy for $S/N=30$, the mass and distance measurement errors are $0.1\%$ and $10\%$ respectively \citep{hg11}. 

Interestingly, the GW signal of IMBH--NCO binaries may carry information on the SMBH 
as well. First, the GWs are modulated by Doppler shift as the binary 
orbits around the SMBH \citep{ymt11}. Furthermore, the SMBH can secularly
excite the eccentricity of the IMBH--NCO binary\citep{ap12,nkly12}. Since the GW signal is extremely sensitive to eccentricity \citep{pm63}, it is likely that this perturbation
would allow us to constrain the parameters of the central SMBH. This effect is the inverse
of that mentioned at the end of Section~\ref{s:GW1}. LIGO or the Einstein Telescope
may measure the SMBH through its influence on the IMBH-NCO system,
while LISA may measure the effects of an NCO through the pertubations of the
SMBH--IMBH signal.

\subsection{NCO--NCO scattering and capture}
Finally, the NCOs which form or become captured by the accretion disk may undergo close encounters and form hard binary systems through GW emission. 
Hard binaries are likely IMBH seeds in AGN disks (see Paper I). For these systems, the GWs are 
detectable with LIGO during pericenter passage \citep{kgm06,okl09,kl12,sam13,east13}.
Similar to the IMBH--NCO case, the signal initially consists of repeated bursts, but 
transitions to a continuous eccentric inspiral in a much shorter time of minutes to days
depending on the impact parameter
(see Eq.~\ref{eq:gravrad} with $\mu\sim M\sim 10\,M_{\odot}$). 

These sources are in the LIGO frequency band before merger even in the circular case. 
For circular NS--NS inspirals, the source localization of the LIGO--VIRGO network is
very poor $50\;\mathrm{deg}$ or more for $S/N \sim 15$, but it improves to within than $10\;\mathrm{deg}$ with future extensions
of the network through KAGRA\footnote{\url{http://gw.icrr.u-tokyo.ac.jp/lcgt/}} or LIGO-India\footnote{\url{http://www.gw-indigo.org/tiki-index.php?page=LIGO-India}} \citep{n10,n11,veitch12}. 
Similar to the IMBH--NCO case, the source localization for eccentric sources is 
expected to be much better. The pertubations caused by the IMBH and the SMBH 
may be detectable. Remarkably, NS/BH and NS/NS binaries may have coincident electromagnetic 
counterparts, i.e. short-hard gamma ray bursts. The coincidence in time of an electromagnetic 
signal like a gamma-ray burst or the FeK$\alpha$ ripple effect allows us to locate the likely 
GW source.

\section{Conclusions}
\label{sec:conclusions}
If Intermediate mass black holes (IMBHs) can grow efficiently in AGN disks, the AGN host  should exhibit myriad observational signatures. IMBHs that open gaps in AGN disks will exhibit strong observational parallels with gapped protoplanetary disks and may be detectable near merger in the broad FeK$\alpha$ line. LINER activity may be due to a weakly accreting MBH binary in a large disk cavity. If IMBHs do not open a gap, detection depends on signatures of accretion onto the IMBH (including tidal disruption events). We summarize observational signatures and compare them to current data where possible or suggest future observations.

\section*{Acknowledgements}
We thank the referee for a report that helped us condense and focus this paper. We acknowledge very useful discussions with Tahir Yaqoob, Stephan Rosswog, Zoltan Haiman, Ari Laor, Hagai Perets, Kayhan G\"{u}ltekin and Mordecai Mac Low. BM and KESF acknowledge support from NASA-APRA08-0117 and NSF PAARE AST-1153335. BK was supported in part by the W.M. Keck Foundation Fund of the Institute for Advanced Study and NASA grant NNX11AF29G. WL acknowledges support by the National Science Foundation under grant No. AST10-09802. This work was performed in part under contract with the California Institute of Technology (Caltech) funded by NASA through the Sagan Fellowship Program executed by the NASA Exoplanet Science Institute.


\label{lastpage}

\end{document}